\documentclass[12pt]{article}
\usepackage{graphicx,epsfig,cite,amsmath,amssymb}
\usepackage{a4wide}
\usepackage{cite}
\usepackage{hyperref}

\newcommand{\be}{\begin{equation}}
\newcommand{\ee}{\end{equation}}
\newcommand{\ba}{\begin{eqnarray}}
\newcommand{\ea}{\end{eqnarray}}

\newcommand{\cO}{{\cal O}}

\newcommand{\gevs}{~\mbox{GeV}^2}

\newcommand{\leff}{\ensuremath{\text{L}_{10}}^{\rm eff}}  
\newcommand{\ceff}{\ensuremath{\text{C}_{87}}^{\rm eff}}  

\begin{document}
\begin{titlepage}
\begin{flushright}
{CAFPE-139/10}\\  {FTUV/10-0428} \\ IFIC/10-10 \\
{UG-FT-269/10}
\end{flushright}
\vspace{2cm}
\begin{center}
{\large\bf Pinched weights and Duality Violation in QCD Sum Rules:\\ a critical analysis}
\vfill
{\bf Mart\'{\i}n Gonz\'alez-Alonso$^a$, Antonio Pich$^a$ and Joaquim Prades$^b$}\\[0.5cm]
${}^a$
Departament de F\'{\i}sica Te\`orica and IFIC, Universitat de Val\`encia-CSIC,\\
Apt. Correus 22085, E-46071 Val\`encia, Spain\\
${}^b$ CAFPE and Departamento de F\'{\i}sica Te\'orica y del Cosmos,\\
Universidad de Granada, Campus de Fuente Nueva, E-18002 Granada, Spain\\[0.5cm]
\end{center}
\vfill
\begin{abstract}
We analyze the so-called pinched weights, that are generally thought to reduce the violation of quark-hadron duality in Finite-Energy Sum Rules. After showing how this is not true in general, we explain how to address this question for the LR correlator and any particular pinched weight taking advantage of our previous work~\cite{GonzalezAlonso:2010rn}, where the possible high-energy behavior of the LR spectral function was studied
. In particular we show that the use of pinched weights allows to determine with high accuracy the dimension six and eight contributions in the operator product expansion, $\cO_6= \left(-4.3^{+0.9}_{-0.7}\right)\cdot 10^{-3}~\mbox{GeV}^{6}$ and $\cO_8= \left(-7.2^{+4.2}_{-5.3}\right)\cdot 10^{-3}~\mbox{GeV}^{8}$.
\end{abstract}

\vfill

\end{titlepage}

\section{Introduction}

In a recent work \cite{GonzalezAlonso:2010rn} we have analyzed the violation of quark-hadron duality (DV) of a given QCD Sum Rule with the non-strange LR correlator $\Pi(q^2) \equiv \Pi_{ud,LR}^{(0+1)}(q^2)$ defined by
\ba
\label{eq:LRcorrelator2}
 \Pi^{\mu\nu}_{ud,LR}(s)\,
&=&\,  i \int \mathrm{d}^4 x \; \mathrm{e}^{i q x} \, \langle 0 | T \left( L_{ud}^\mu(x) R_{ud}^\nu(0)^\dagger \right) | 0 \rangle \nonumber \\
&=& (-g^{\mu\nu} q^2 + q^\mu q^\nu )~\Pi^{(0+1)}_{ud,LR}(q^2) + g^{\mu\nu} q^2~\Pi^{(0)}_{ud,LR}(q^2) \, ,
\ea
where $L_{ud}^\mu(x)\equiv \overline{u} \gamma^\mu (1-\gamma_5) d$ and $R_{ud}^\mu(x)\equiv \overline{u} \gamma^\mu (1+\gamma_5) d$.

A QCD Sum Rule \cite{Shifman:1978bx} takes advantage of the analytic properties of the correlator to relate its imaginary part in the positive real $q^2$-axis (where hadrons lie) with its value in the rest of the complex plane, where the Operator-Product Expansion (OPE) allows us to calculate it in terms of quarks and gluons:
$\Pi^{\rm{OPE}}(s) = \sum_k \mathcal{O}_{2k}/(-s)^k$. 
The DV comes from the fact that this OPE breaks down in the vicinity of the positive real $q^2$-axis. We can write a general QCD Sum Rule for the LR correlator in the following form 
\ba
\label{eq:FESRwithDV}
\lefteqn{
\int^{s_0}_{s_{\rm th}} \!\mathrm{d}s\; w(s) \, \rho(s) ~ + ~ \frac{1}{2 \pi i} \, \oint_{|s|=s_0} \!\!\!\! \mathrm{d}s\; w(s) \,\Pi^{\rm{OPE}}(s) ~ + ~ \mathrm{DV}[w(s),s_0]}
&& \nonumber \\ && \hskip 7cm
=\; 2 f_\pi^2\, w(m_\pi^2) ~ + ~ \underset{s=0}{\text{Res}} \left[ w(s) \, \Pi(s)\right] ,\quad
\ea
where $\rho(s)\equiv\frac{1}{\pi}\,\mathrm{Im}\Pi(s)$ and w(s) is an arbitrary weight function that is analytic in the whole complex plane except in the origin (where it can have poles). The violation of quark-hadron duality is formally defined as 
\cite{Shifman:2000jv,Cirigliano:2002jy,Cirigliano:2003kc,Cata:2005zj,GON07,GonzalezAlonso:2010rn}
%
\ba
\label{eq:DV1}
{\rm DV} [w(s),s_0] ~\equiv~ \frac{1}{2 \pi i} \, \oint_{|s|=s_0} \mathrm{d}s\; w(s) \left( \Pi(s) - \Pi^{\rm{OPE}}(s)\right)~.
\ea
Using analyticity one can write the DV in the following form \cite{Shifman:2000jv,Chibisov:1996wf,Cata:2005zj,GON07}
\ba
\label{eq:DV2}
{\rm DV} [w(s),s_0]  ~=~ \int^{\infty}_{s_0} \mathrm{d}s~w(s)~\rho(s)~,
\ea
that shows how the DV is nothing but the part of the integral of the spectral function 
 that we are not including in the sum rule. In Ref. \cite{GonzalezAlonso:2010rn} we have studied the DV from this perspective, using the following parametrization
\ba
\label{eq:model}
\rho(s\ge s_z) &=& \kappa~ e^{-\gamma s} \sin(\beta (s-s_z))~,
\ea
for the spectral function beyond $s_z \sim 2.1\gevs$ and finding the region in the 4-dimensional parameter space that is compatible with the most recent experimental data \cite{Schael:2005am} and the following theoretical constraints: first and second Weinberg Sum Rules \cite{Weinberg:1967kj} (WSRs) and the sum rule of Das et al. \cite{Das:1967it} that gives the electromagnetic mass difference of pions ($\pi$SR). 
The parametrization (\ref{eq:model})
emerges naturally in a resonance-based model \cite{Blok:1997hs,Shifman:1998rb,Shifman:2000jv} that has been used recently 
to study the violation of quark-hadron duality \cite{Cata:2005zj,GON07,Cata:2008ye,Cata:2008ru}, although without imposing the previously explained theoretical constraints in the numerical analysis.

In Ref. \cite{GonzalezAlonso:2010rn} we used this parametrization to calculate the DV associated to Finite-Energy Sum Rules (FESRs) with the weights $w(s)=s^n$ ($n=-2,-1,+2,+3$), but it can be used to analyze any other QCD Sum Rule with the LR correlator. In this letter we would like to apply the results of \cite{GonzalezAlonso:2010rn} to the so-called pinched-weight FESRs, where the standard weight $s^n$ is substituted by a polynomial weight that vanishes at $s=s_0$ (or near this point).

It has been often assumed that the use of pinched-weights (PWs) minimizes the DV\footnote{It must be emphasized that the PW functions are also useful because they are expected to minimize the experimental errors, since they suppress the region near the kinematical end point.} \cite{LeDiberder:1992te,LeDiberder:1992fr,Dominguez:1998wy,Cirigliano:2002jy,Dominguez:2003dr,Cirigliano:2003kc,Bordes:2005wv,Dominguez:2006ct,Maltman:2008nf,Maltman:2008ib}, since they suppress the contribution from the most problematic region in the contour integral of Eq.~\eqref{eq:DV1}, close to the real axis  \cite{Poggio:1975af}. However the alternative expression for the DV given in Eq.~\eqref{eq:DV2} shows that things are more subtle \cite{Cata:2005zj,GON07,GonzalezAlonso:2010rn} and that the assumption is not necessarily true, since a PW function will indeed suppress the first part of this hadronic integral but at the same time may enhance the high-energy tail that can become important. If the final balance is positive and the weight function does its job minimizing the DV contribution 
 is something that depends on the particular weight used and on how fast the spectral function goes to zero, something that is not known theoretically.

This question about the convenience of the use of these PWs is very entangled with the more general question of how to estimate the duality violation of a given sum rule. The observation of a more stable plateau in the final part of the data range is the standard requirement to check if the weight improves the situation, and the deviations from the plateau the standard way of estimating the remaining DV. However it is important to notice that the existence of the plateau is a necessary but not sufficient condition, because it could be temporary. This is particularly plausible because the PWs produce curves that have derivative zero in the second duality point\footnote{The duality points are two particular points (located at $s_0\sim1.5\gevs$ and $s_0\sim2.6\gevs$) where both the first and second WSRs happen to be satisfied, i.e. where their DV contributions vanish.} ($s_0\sim2.6\gevs$), which is very near of the end of the data. That is, they produce a fake plateau, that can induce to the possibly wrong conclusion that the DV is negligible for that weight and that value of $s_0$.

Here fake means that the correlations between the experimental points of the plateau are extremely high and such that we do not have several points indicating the same value, but just one point drawn several times. In principle, a fit of these points to a straight line is sensitive to the correlations and would tell us if the plateau is real or it has been artificially created by the weight function, but in practice this is not always possible, since the high correlations among points prevent us from using the standard $\chi^2$-fit, as explained in \cite{D'Agostini:1993uj}\footnote{This situation was found e.g. in the determination of the V-A condensates in Ref. \cite{Cirigliano:2002jy,Cirigliano:2003kc}.}.


The results obtained in our previous analysis \cite{GonzalezAlonso:2010rn} allow to address these issues in a quantitative way. In particular, we will study the PW versions of the QCD Sum Rules of Ref.~\cite{GonzalezAlonso:2010rn}, i.e. those that give an estimation of the hadronic parameters $\ceff,\leff,\cO_6$ and $\cO_8$.

\section{Numerical analysis}

We are interested in PW functions that do not introduce new unknown quantities (condensates of higher dimension), since in that case a clean analysis is not possible anymore, and more specifically we will work with pinched-weights $w(s)$ that have a double zero in $s=s_{pw}$, that is
\ba
\label{eq:C87pw}
&&\int^{s_z}_{s_{\rm th}} \!\mathrm{d}s\; \frac{\rho(s)}{s^2}\left( 1-\frac{s}{s_{pw}}\right)^2 \left(  1+\frac{2s}{s_{pw}}\right)
= ~16~C_{87}^{\rm eff} - 6 \frac{f_\pi^2}{s_{pw}^2}+ 4 \frac{f_\pi^2 m_\pi^2}{s_{pw}^3} -\mbox{DV}[w_{-2},s_z]\,, \\
\label{eq:L10pw}
&&\int^{s_z}_{s_{\rm th}} \!\mathrm{d}s~\frac{\rho(s)}{s}\left( 1-\frac{s}{s_{pw}}\right)^2
= -8 L_{10}^{\rm eff} - 4 \frac{f_\pi^2}{s_{pw}}+  2\frac{f_\pi^2 m_\pi^2}{s_{pw}^2}  - \mbox{DV}[w_{-1},s_z]\,, \\
\label{eq:M2pw}
&&\int^{s_z}_{s_{\rm th}} \!\mathrm{d}s~  \rho(s)\left(s - s_{pw} \right)^2
~=\, 2 f_\pi^2 s_{pw}^2 - 4 f_\pi^2 m_\pi^2 s_{pw}~+~2 f_\pi^2 m_\pi^4~+~\cO_6 \, - \mbox{DV}[w_2,s_z]\,, \\
\label{eq:M3pw}
&&\int^{s_z}_{s_{\rm th}} \!\mathrm{d}s~\rho(s) \left(s - s_{pw} \right)^2 \left(s +2 s_{pw} \right)\nonumber\\
&&~~~~~~~~~~~~~~~~~~~~~~~~~~~~~~= -6 f_\pi^2 m_\pi^2 s_{pw}^2 + 4f_\pi^2 s_{pw}^3 + 2 f_\pi^2 m_\pi^6\, -\, \cO_8\, - \mbox{DV}[w_3,s_z]\,.
\ea
%

The results depend on the point $s_{pw}$ where the weight is pinched. In order to suppress the experimental error it is convenient to pinch the weight at the left of the matching point $s_z$, whereas in order to suppress the DV-error (dispersion of the histograms) it is convenient to pinch it at the right of $s_z$. We have scanned the region finding that the optimal choice of $s_{pw}$, that is, where the errors are minimized\footnote{Obviously the optimal point is different for every sum rule \eqref{eq:C87pw} - \eqref{eq:M3pw}, but the differences are negligible within errors.}, is $s_{pw}\sim s_z \sim 2.1\gevs$.

A careful comparison between these PWs and the standard weights $s^n$ shows that in the case of the condensates the former are smaller (in absolute value) than the later  for any $s\ge s_{pw}$, and therefore are expected to generate a smaller DV\footnote{Notice that this is not a mathematical statement, but only a hand-waving estimate and can be altered due to accidental cancellations.}, although the question of how large is the remaining DV is not clear at all. In the case of the chiral parameters $\leff$ and $\ceff$, the convenience of the PW is not known a priori and it depends essentially on how fast the spectral function goes to zero, in order to suppress the enhancement that the PWs produce in the high-energy region (see Eq.~\eqref{eq:DV2}). In other words, the key point is the value of the $\gamma$ parameter, that is around one \cite{GonzalezAlonso:2010rn}. We will show that this value is large enough to suppress the high-energy tail and so to benefit from the use of the PW.

\begin{figure}[t!]
\vfill
\centerline{
\begin{minipage}[t]{.35\linewidth}\centering
\centerline{\includegraphics[width=6.4cm]{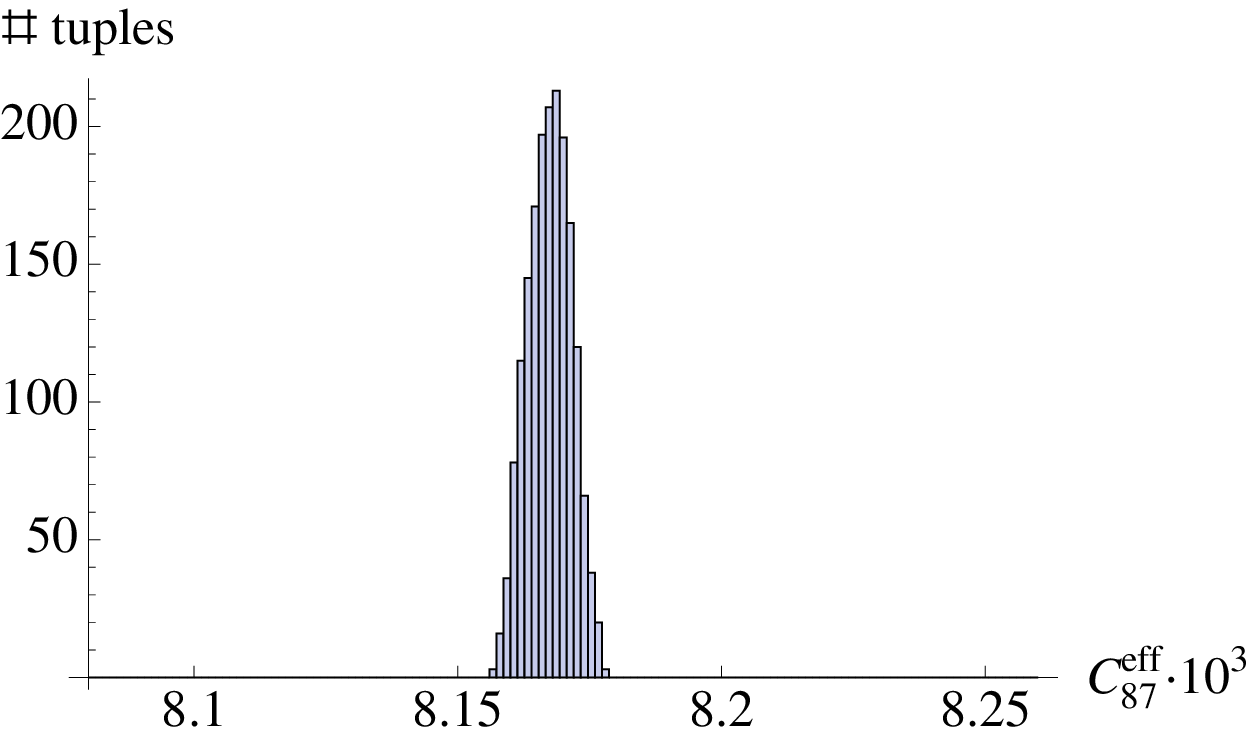}}
\end{minipage}
\hspace{2cm}
\begin{minipage}[t]{.35\linewidth}\centering
\centerline{\includegraphics[width=6.4cm]{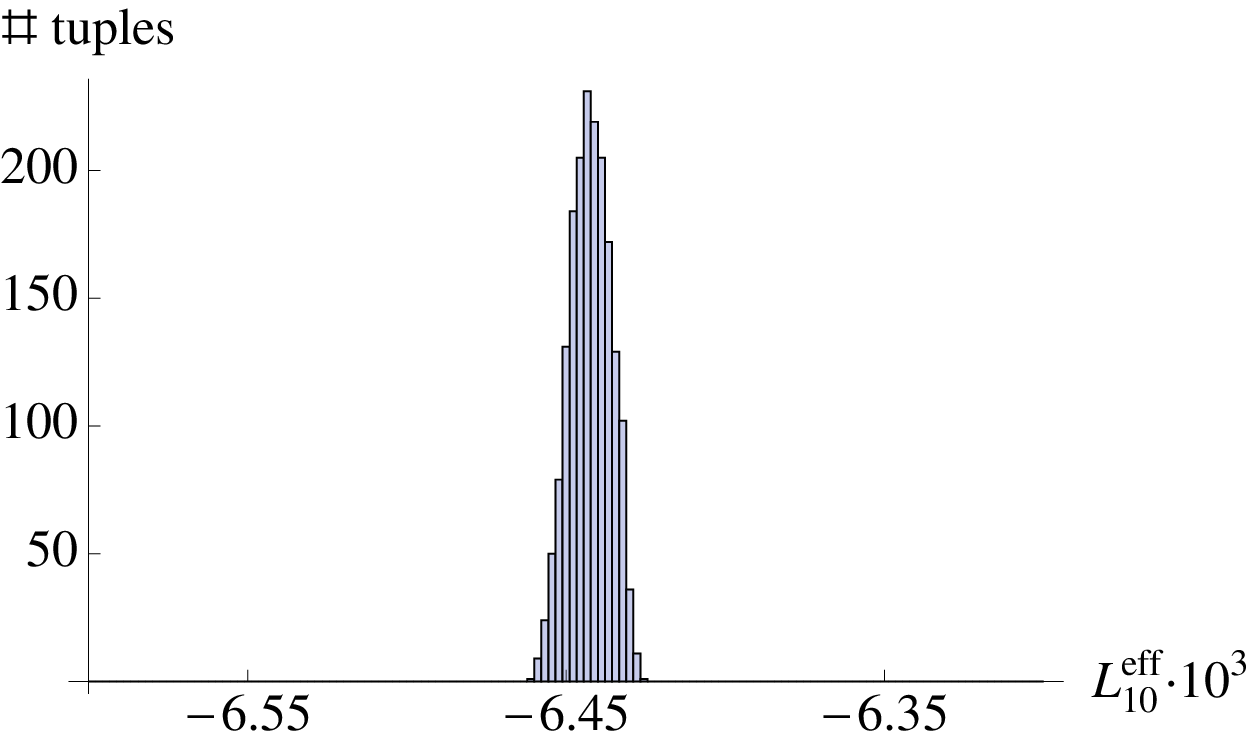}}
\end{minipage}
}
\vspace{0.7cm}
\centerline{
\begin{minipage}[t]{.35\linewidth}\centering
\centerline{\includegraphics[width=6.4cm]{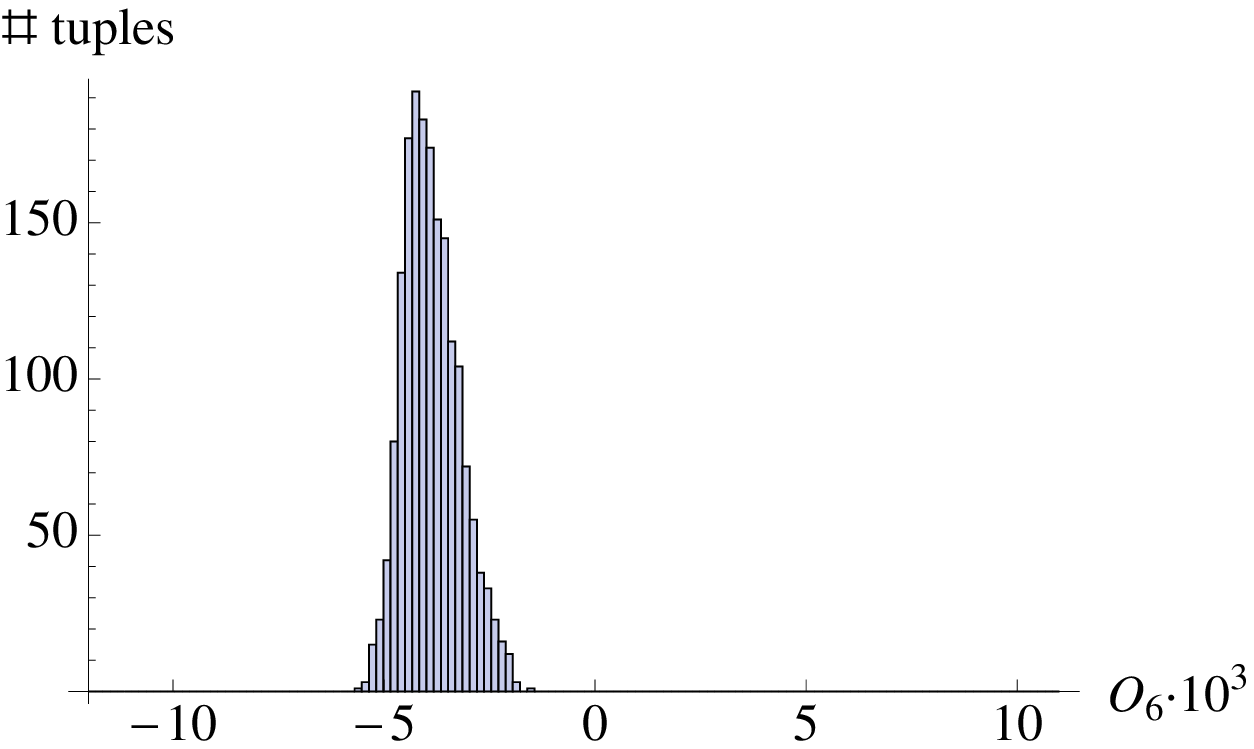}}
\end{minipage}
\hspace{2cm}
\begin{minipage}[t]{.35\linewidth}\centering
\centerline{\includegraphics[width=6.4cm]{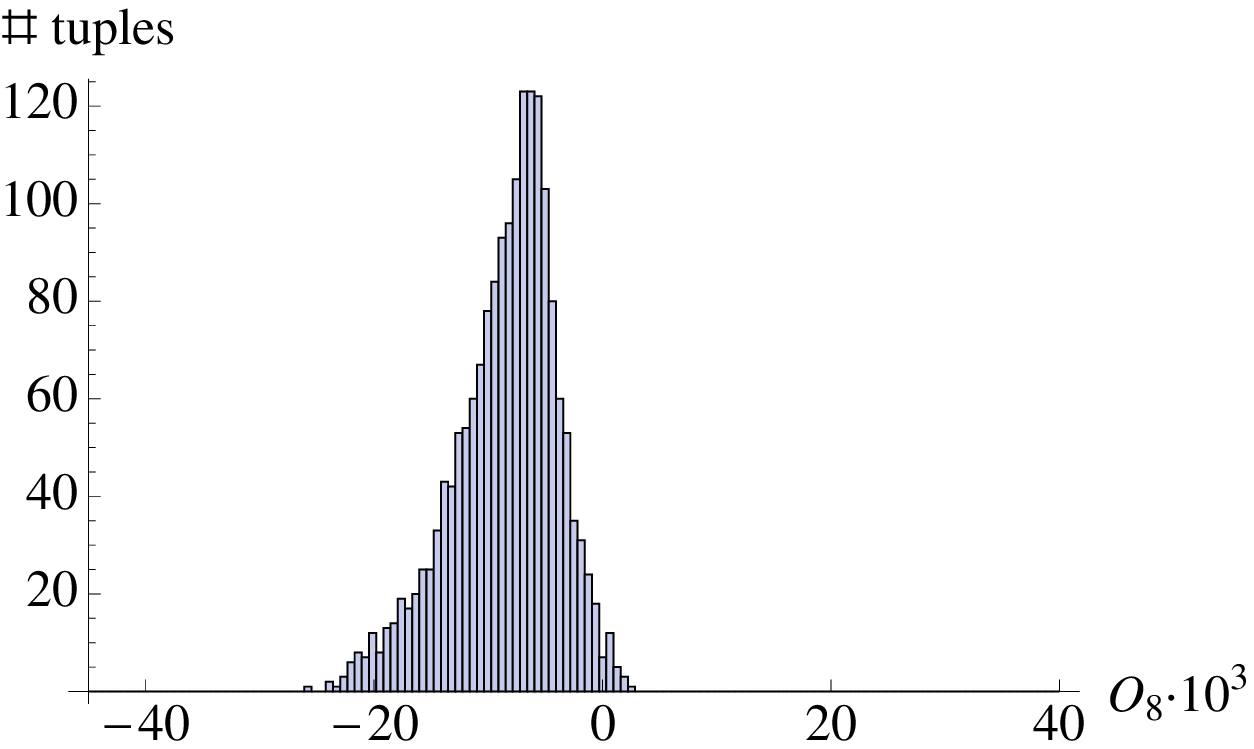}}
\end{minipage}
}
\vfill
\caption{Statistical distribution of values of $C_{87}^{\mathrm{eff}}$ (upper-left), $L_{10}^{\mathrm{eff}}$ (upper-right), $\cO_6$ (lower-left)
and $\cO_8$ (lower-right) for the accepted spectral functions, using the pinched-weight sum rules \eqref{eq:C87pw} - \eqref{eq:M3pw} with $s_{pw}=s_z\sim 2.1\gevs$. The parameters are expressed in GeV to the corresponding power.}
\label{fig:observablesPW2}
\end{figure}

In Ref.~\cite{GonzalezAlonso:2010rn} we have used the parametrization \eqref{eq:model} for the spectral function $\rho(s)$ and we have analyzed the allowed parameter space once the experimental and theoretical constraints are taken into account. In other words, we have generated a large number of ``acceptable'' spectral functions, compatible with both QCD and the data.
The differences among them determine how much freedom is left for the behavior of the spectral function beyond the kinematical end of the $\tau$ data. In particular we can calculate the value of the parameters $\ceff,\leff,\cO_6$ and $\cO_8$ obtained through the sum rules \eqref{eq:C87pw} - \eqref{eq:M3pw} for each of these possible spectral functions. The results of this process are given in Fig. \ref{fig:observablesPW2}, which shows the statistical distribution of the generated values. We can see that the histograms are much more peaked around their central values than those obtained in Ref.~\cite{GonzalezAlonso:2010rn} with standard weights.

Let us remind that in addition to the error associated to the DV (estimated from the dispersion of the histograms) we have the experimental ALEPH error, and both depend on the used weight function. In principle one expects the PWs to minimize also the experimental uncertainties, since they suppress the region near the kinematical end point\footnote{Notice below that in the case of $\cO_8$ this does not happen. This is because the PW enhances the low-energy region errors sizably.}.

The associated numerical results are (we give the $68\%$ probability region).
\ba
\label{eq:C87resultPW68}
\ceff &=&\left(8.168\, {}^{+0.003}_{-0.004}\pm\!0.12 \right)\cdot\! 10^{-3}~\mbox{GeV}^{-2}\; =\; \left(8.17\pm 0.12 \right)\cdot\! 10^{-3}~\mbox{GeV}^{-2},\\
\label{eq:L10resultPW68}
\leff &=& \left(-6.444\, {}^{+0.007}_{-0.004}\pm0.05 \right)\cdot 10^{-3}\; =\;  \left(-6.44 \pm 0.05 \right)\cdot 10^{-3}~,\\
\label{eq:O6resultPW68}
\cO_6&=& \left(-4.33\, {}^{+0.68}_{-0.34}\pm0.65 \right)\cdot 10^{-3}~\mbox{GeV}^{6}\; =\;  \left(-4.3\, {}^{+0.9}_{-0.7}\right)\cdot 10^{-3}~\mbox{GeV}^{6}~,\\
\label{eq:O8resultPW68}
\cO_8&=& \left(-7.2\, {}^{+3.1}_{-4.4}\pm 2.9\right)\cdot 10^{-3}~\mbox{GeV}^{8}\; =\;  \left(-7.2\, {}^{+4.2}_{-5.3}\right)\cdot 10^{-3}~\mbox{GeV}^{8}~,
\ea
where the first error is that associated to the high-energy region (integral from $s_z$ to infinity), that we compute from the dispersion of the histograms of Fig. \ref{fig:observablesPW2}, and the second error is that associated to the low-energy region (integral from zero to $s_z$), that we compute in a standard way from the ALEPH data. For the sake of comparison we show the analogous results obtained in Ref.~\cite{GonzalezAlonso:2010rn}: $C_{87}^{\mathrm{eff}} = \left(8.17\,\pm0.12\right)\cdot 10^{-3}~\mbox{GeV}^{-2}$, $L_{10}^{\mathrm{eff}}=\left(-6.46\, {}^{+\, 0.08}_{-\, 0.07}\right)\cdot 10^{-3}$, $\cO_6=\left(-5.4\, {}^{+\, 3.6}_{-\, 1.6}\right)\cdot 10^{-3}~\mbox{GeV}^6$ and $\cO_8 = \left(-8.9\, {}^{+\, 12.6}_{-\, 7.4}\right)\cdot 10^{-3}~\mbox{GeV}^8$, where we can clearly see the improvement achieved with the PWs.

Since the first error in Eqs.~\eqref{eq:C87resultPW68}-\eqref{eq:O8resultPW68} is not Gaussian we show also here the $95\%$ probability results:
\ba
\ceff &=& \left(8.168\, {}^{+0.005}_{-0.008}\pm\!0.24 \right)\cdot\! 10^{-3}~\mbox{GeV}^{-2}\; =\;  \left(8.17 \pm\! 0.24\right)\cdot \!10^{-3}~\mbox{GeV}^{-2},\\
\leff &=& \left(-6.444\, {}^{+0.011}_{-0.011}\pm0.1 \right)\cdot 10^{-3}\; =\;  \left(-6.4 \pm 0.1\right)\cdot 10^{-3}~,\\
\cO_6&=& \left(-4.33\, {}^{+1.70}_{-0.68}\pm1.3 \right)\cdot 10^{-3}~\mbox{GeV}^{6}\; =\;  \left(-4.3\, {}^{+2.1}_{-1.5}\right)\cdot 10^{-3}~\mbox{GeV}^{6}~,\\
\label{eq:O8resultPW}
\cO_8&=& \left(-7.2\, {}^{+6.3}_{-11.3}\pm 5.8\right)\cdot 10^{-3}~\mbox{GeV}^{8}\; =\;  \left(-7.2\, {}^{+8.6}_{-12.7}\right)\cdot 10^{-3}~\mbox{GeV}^{8}~.
\ea

\section{Beyond the dimension eight condensate}
\label{sec:O10}

\begin{figure}[t!]
\vfill
\centerline{
\begin{minipage}[t]{.35\linewidth}\centering
\centerline{\includegraphics[width=6.4cm]{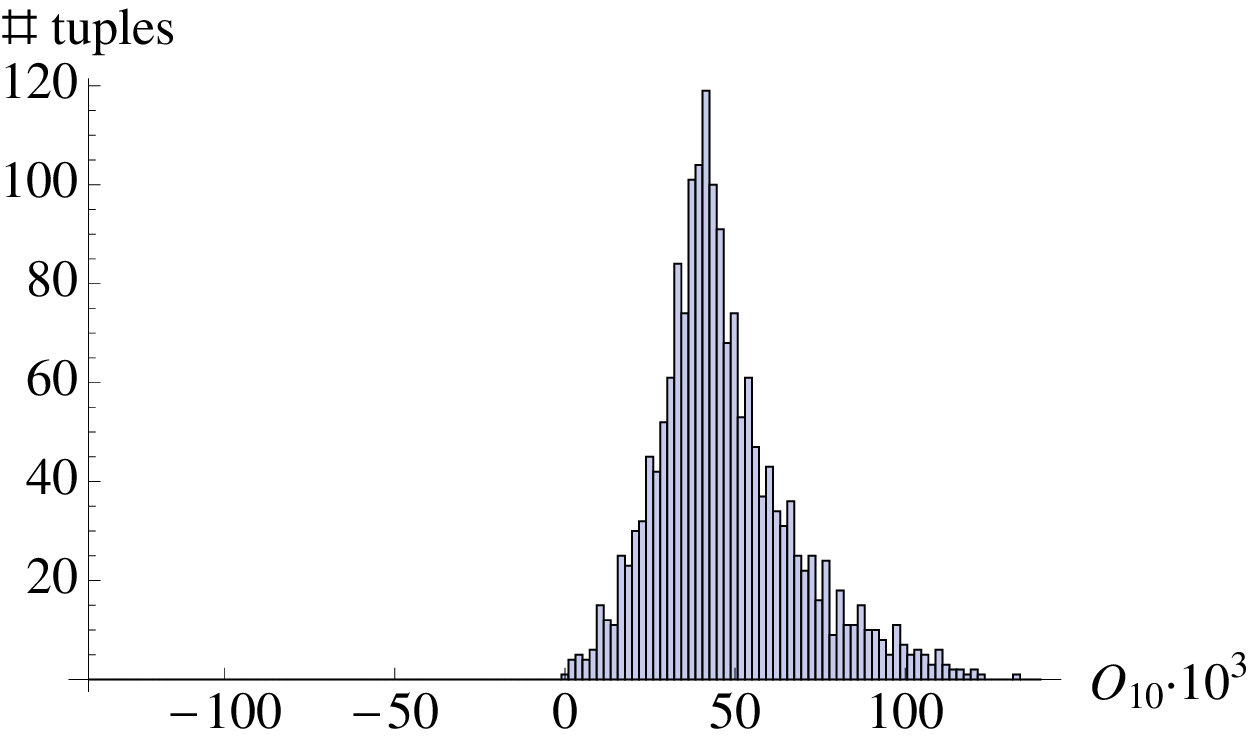}}
\end{minipage}
\hspace{2cm}
\begin{minipage}[t]{.35\linewidth}\centering
\centerline{\includegraphics[width=6.4cm]{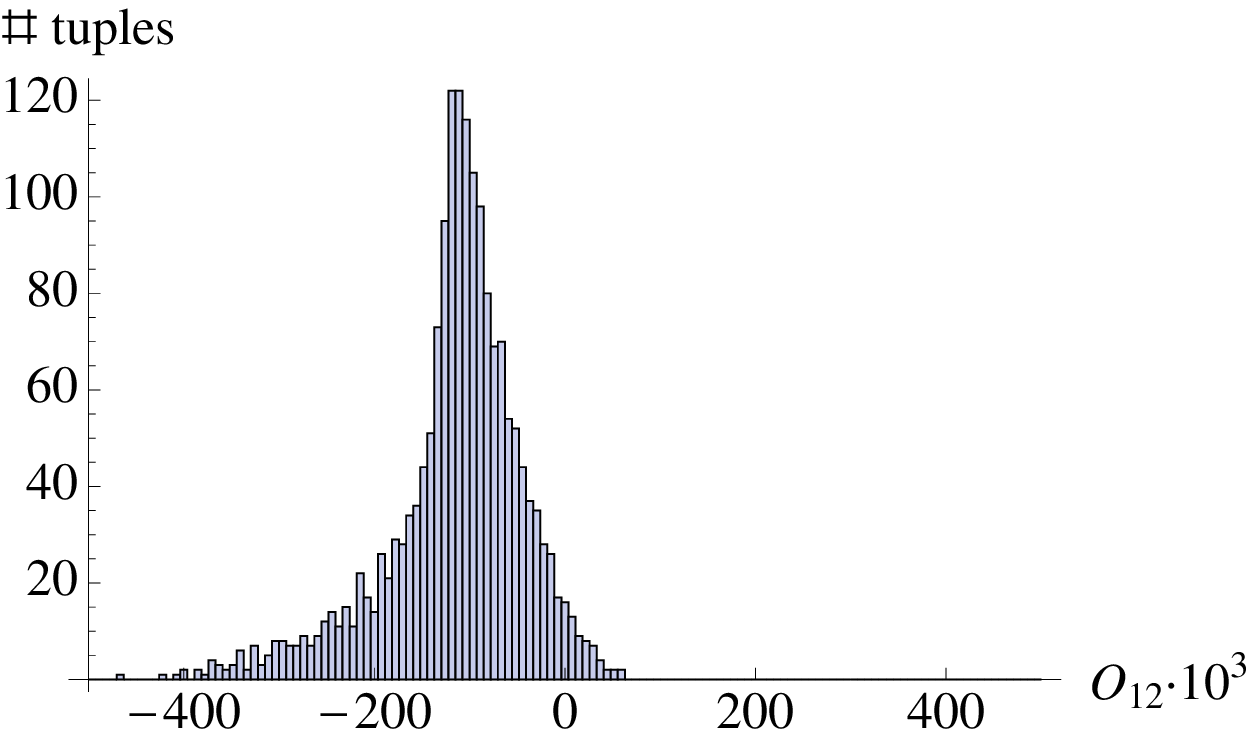}}
\end{minipage}
}
\vspace{0.7cm}
\centerline{
\begin{minipage}[t]{.35\linewidth}\centering
\centerline{\includegraphics[width=6.4cm]{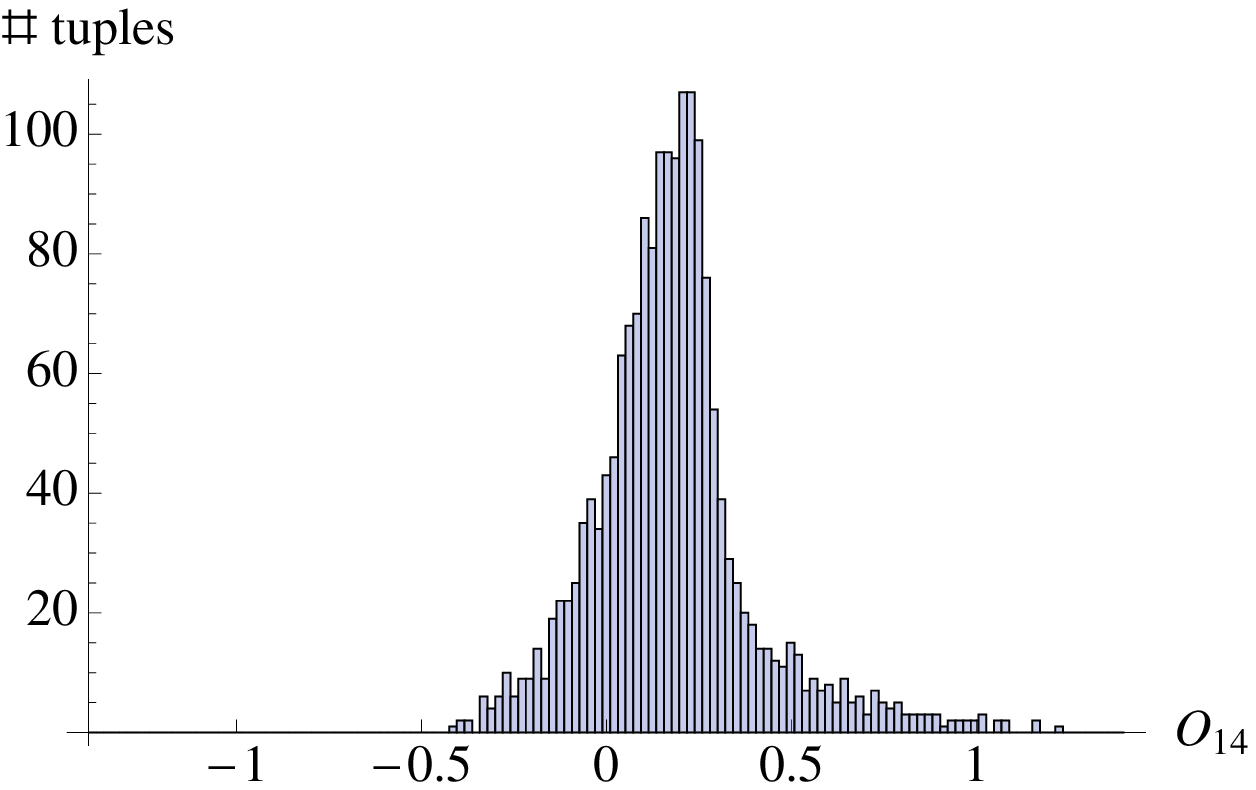}}
\end{minipage}
\hspace{2cm}
\begin{minipage}[t]{.35\linewidth}\centering
\centerline{\includegraphics[width=6.4cm]{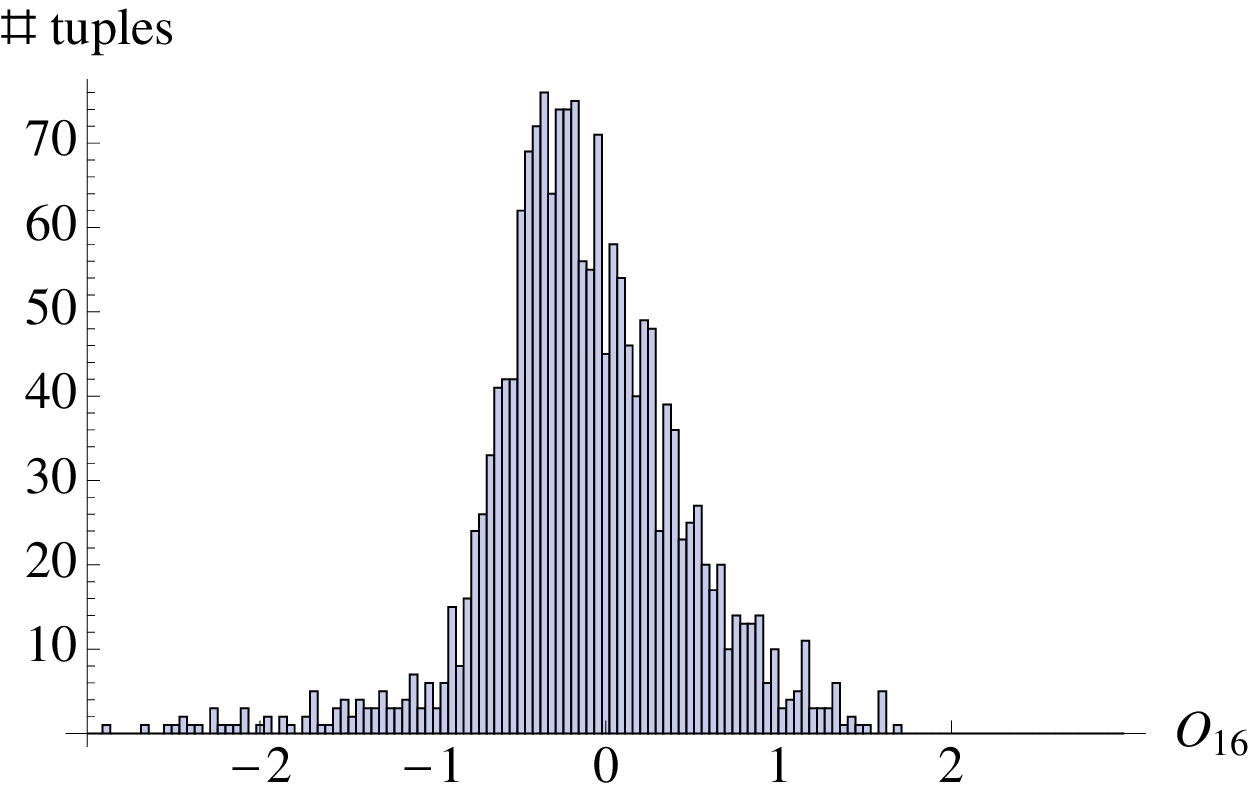}}
\end{minipage}
}
\vfill
\caption{Statistical distribution of values of $\cO_{10,12,14,16}$ for the accepted spectral functions, using the PW sum rules \eqref{eq:M4pw} - \eqref{eq:M7pw} with $s_{pw}=s_z\sim 2.1\gevs$. The parameters are expressed in GeV to the corresponding power.}
\label{fig:highercond}
\end{figure}

We can play the same game with higher-dimensional condensates, where using again pinched weights $w(s)$ that have a double zero in $s=s_{pw}$ we have
\ba
\label{eq:M4pw}
&& \int^{s_z}_{s_{\rm th}} \mathrm{d}s~\rho(s) ~\left(s - s_{pw} \right)^2 \left(s^2 +2 s_{pw}s + 3s_{pw}^2 \right) \nonumber\\
&&~~~~~~~~~~~~~~~~~~~~= -8 f_\pi^2 m_\pi^2 s_{pw}^3 + 6f_\pi^2 s_{pw}^4 + 2 f_\pi^2 m_\pi^8+\, \cO_{10}\, - ~\mbox{DV}[w_4,s_z]~,\\
\label{eq:M5pw}
&& \int^{s_z}_{s_{\rm th}} \mathrm{d}s~\rho(s) ~\left(s - s_{pw} \right)^2 \left(s^3 +2 s_{pw}s^2 + 3s_{pw}^2s + 4s_{pw}^3\right) \nonumber\\
&&~~~~~~~~~~~~~~~~~~~~= -10 f_\pi^2 m_\pi^2 s_{pw}^4 + 8f_\pi^2 s_{pw}^5 + 2 f_\pi^2 m_\pi^{10}-\, \cO_{12}\, - ~\mbox{DV}[w_5,s_z]~,\\
\label{eq:M6pw}
&& \int^{s_z}_{s_{\rm th}} \mathrm{d}s~\rho(s) ~\left(s - s_{pw} \right)^2 \left(s^4 +2 s_{pw}s^3 + 3s_{pw}^2s^2 + 4s_{pw}^3s + 5s_{pw}^4\right) \nonumber\\
&&~~~~~~~~~~~~~~~~~~~~= -12 f_\pi^2 m_\pi^2 s_{pw}^5 + 10f_\pi^2 s_{pw}^6 + 2 f_\pi^2 m_\pi^{12}+\, \cO_{14}\, - ~\mbox{DV}[w_6,s_z]~,\\
\label{eq:M7pw}
&& \int^{s_z}_{s_{\rm th}} \mathrm{d}s~\rho(s) ~\left(s - s_{pw} \right)^2 \left(s^5 +2 s_{pw}s^4 + 3s_{pw}^2s^3 + 4s_{pw}^3s^2 + 5s_{pw}^4s+6s_{pw}^5\right) \nonumber\\
&&~~~~~~~~~~~~~~~~~~~~= -14 f_\pi^2 m_\pi^2 s_{pw}^6 + 12f_\pi^2 s_{pw}^7 + 2 f_\pi^2 m_\pi^{14}-\, \cO_{16}\, - ~\mbox{DV}[w_7,s_z]~.
\ea
Working again with $s_{pw}\sim s_z \sim 2.1\gevs$ we find the results shown in Fig. \ref{fig:highercond}.
 The associated numerical values are (68\% C.L.)
\ba
\cO_{10}&=& \left(+4.1\, {}^{+1.8}_{-1.6}\right)\cdot 10^{-2}~\mbox{GeV}^{10}~,\\
\cO_{12}&=& \left(-0.12\, {}^{+0.07}_{-0.03} \right) ~\mbox{GeV}^{12}~,\\
\cO_{14}&=& \left(+0.2\, {}^{+0.1}_{-0.2} \right)~\mbox{GeV}^{14}~,\\
\cO_{16}&=& \left(-0.2\, {}^{+0.5}_{-0.4}\right)~\mbox{GeV}^{16}~,
\ea
where all the errors come from the dispersion of our histograms since the experimental error is very much smaller for these higher-dimensional condensates. The $95\%$ probability results are:
\ba
\cO_{10}&=& \left(+4.1\, {}^{+5.6}_{-3.1}\right)\cdot 10^{-2}~\mbox{GeV}^{10}~,\\
\cO_{12}&=& \left(-0.12\, {}^{+0.13}_{-0.16} \right) ~\mbox{GeV}^{12}~,\\
\cO_{14}&=& \left(+0.2\pm 0.5 \right)~\mbox{GeV}^{14}~,\\
\cO_{16}&=& \left(-0.2\, {}^{+1.8}_{-1.1} \right)~\mbox{GeV}^{16}~.
\ea
It is really impressive that the sign of the condensates can be established for $\cO_{10}$ and $\cO_{12}$ since the importance of the high-energy region in their determination is huge. One could have expected that the differences between our possible spectral functions would generate a huge error in these higher-dimensional condensates, but our conditions (WSRs+$\pi$SR+data) have turned out to be very restrictive about the acceptable spectral functions allowing quite precise extractions.

\section{Comparisons and summary}

We have used the method developed in Ref.~~\cite{GonzalezAlonso:2010rn} to analyze the error of different pinched-weight Finite-Energy Sum Rules and to extract the value of different hadronic parameters. Comparing the results obtained here with those of Ref.~\cite{GonzalezAlonso:2010rn} we see that, as theoretically expected, the use of the pinched weights is less beneficial to the determination of the low-energy constants $\leff$ and $\ceff$ than to the determination of the condensates. Our final results for the former are in excellent agreement with the most precise determination of them~\cite{GonzalezAlonso:2008rf}:
$C_{87}^{\mathrm{eff}} =( 8.18\pm0.14 ) \cdot 10^{-3}~\mbox{GeV}^{-2}$ \ and \
$L_{10}^{\mathrm{eff}} = - ( 6.48\pm0.06 ) \cdot 10^{-3}$.
Notice that, even if these determinations are also based on the PW sum rules \eqref{eq:C87pw} and \eqref{eq:L10pw}, the estimation of the error presented here is obtained through a completely different method, based on more solid grounds and represents a confirmation of them.

\begin{figure}[t]
\vfill
\centerline{
\begin{minipage}[t]{.35\linewidth}\centering
\centerline{\includegraphics[width=7.0cm]{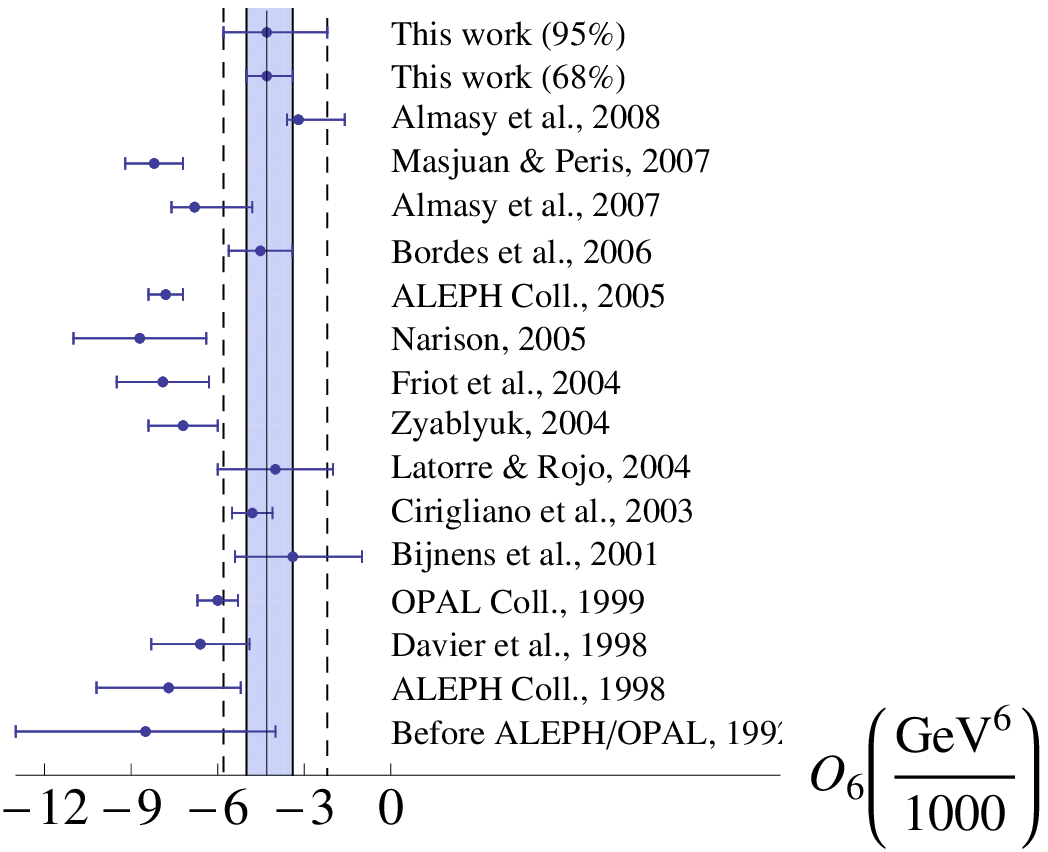}}
\end{minipage}
\hspace{2cm}
\begin{minipage}[t]{.35\linewidth}\centering
\centerline{\includegraphics[width=7.0cm]{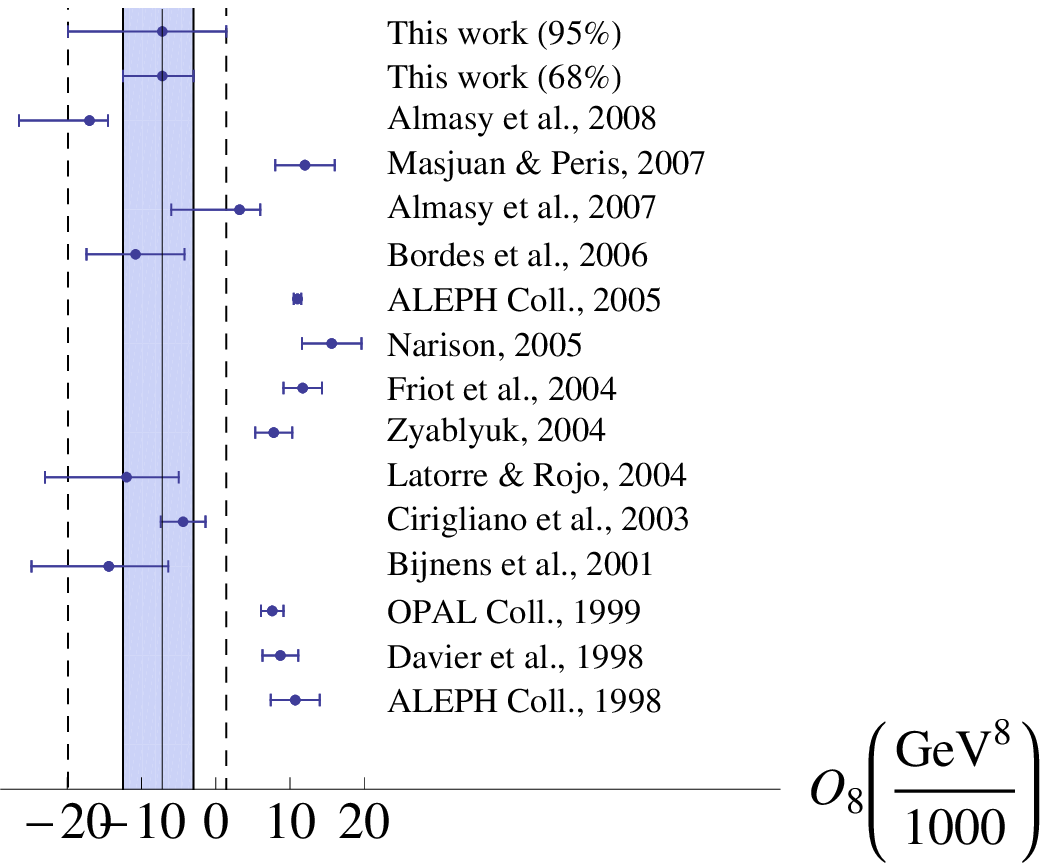}}
\end{minipage}
}
\vfill
\caption[]{Comparison of our results for $\cO_6$ (left) and $\cO_8$ (right) with previous determinations \cite{Schael:2005am,Ackerstaff:1998yj,Barate:1998uf,Davier:1998dz,Bijnens:2001ps,Cirigliano:2002jy,Cirigliano:2003kc,Ioffe:2000ns,Geshkenbein:2001mn,Zyablyuk:2004iu,Rojo:2004iq,Narison:2004vz,Dominguez:2003dr,Bordes:2005wv,Peris:2000tw,Friot:2004ba,Almasy:2006mu,Masjuan:2007ay,Almasy:2008xu} (we show for every method the most recent determination). The blue bands show our results at 65\% C.L., while the 95\% probability regions are indicated by the dotted lines.}
\label{fig:comparisonPW2}
\end{figure}

We have obtained quite precise measurements for the condensates $\cO_6$ and $\cO_8$ using the PW sum rules \eqref{eq:M2pw} and \eqref{eq:M3pw}. In this way we have checked that the PW succeeds in minimizing the errors and we can conclude that the most recent experimental data provided by ALEPH, together with the theoretical constraints (WSRs and $\pi$SR), fix with accuracy the value of $\cO_6$ and almost determine\footnote{One can see in our final result \eqref{eq:O8resultPW} that at $2\sigma$ a positive value of $\cO_8$ is already allowed, but it must not be forgotten that the distribution is highly non-Gaussian and we can see in the corresponding histogram of Fig.~\ref{fig:observablesPW2} that the possibility of being positive is negligible.} the sign of $\cO_8$. Our results are compared in Fig.~\ref{fig:comparisonPW2} with previous determinations of $\cO_6$ and $\cO_8$. One recognizes in the figure the existence of two groups of results that disagree between them. For $\cO_6$ there is a small tension between a bigger or smaller value, whereas in the case of $\cO_8$ the disagreement affects the sign and is more sizable. As can be seen in Table \ref{tab:highdim}, these discrepancies also appear in higher-dimensional condensates, that we have also extracted applying the same method.

\begin{table}[tbh]\centering
	\begin{tabular}{|c|c|c|c|c|}
	\hline
										& $     \cO_{10}\times 10^3$  &      $\cO_{12}\times 10^3$  &   $\cO_{14}\times 10^3$  &    $\cO_{16}\times 10^3$	\\
	\hline
	This work							&	$+41^{+18}_{-16}$	&	$-120^{+70}_{-30}$		&	$+200^{+100}_{-200}$	&	 $-200^{+500}_{-400}$	\\[0.1cm]
	\hline
	Masjuan $\&$ Peris \cite{Masjuan:2007ay}	&	$-14\pm 12$	&								&							&						\\
	Narison \cite{Narison:2004vz}			&	$-17.1\pm 4.4$		&	$+14.7\pm 3.7$			&	$-9.6\pm 3.1$			&	$+4.3\pm1.9$		\\
	Friot et al. \cite{Friot:2004ba}			&	$-13.2\pm 3.6$		&	$+13.3\pm 3.9$			&	$-12.8\pm 3.9$		&	 $+11.9\pm 3.8$	\\
	Zyablyuk \cite{Zyablyuk:2004iu}		&	$-4.5\pm 3.4$			&								&							&						\\
	\hline
	Almasy et al. \cite{Almasy:2008xu}		&	$+66^{+40}_{-14}$&							&							&							\\
	Bordes et al. \cite{Bordes:2005wv}		&	$+72\pm 28$		&	$-240\pm 50$			&							&							\\
	Latorre $\&$ Rojo \cite{Rojo:2004iq}		&	$+78\pm 24$		&	$-260\pm 80$			&							&							 \\
	Cirigliano et al. \cite{Cirigliano:2003kc}	&	$+48 \pm 10$		&	$-160 \pm 30$			&	$+430 \pm 60$		&	$-1030 \pm 140$		\\
	\hline	
	\end{tabular}
\caption{Comparison of our determination of $\cO_{10,12,14,16}$ with other works. The condensates are expressed in GeV to the corresponding power. The results shown for Ref.~\cite{Cirigliano:2003kc} are those obtained with the old ALEPH data (with the OPAL data the numbers are not very different), and the results shown for Ref.~\cite{Friot:2004ba} are those obtained with the minimal hadronic ansatz, that is, without the addition of the $\rho^\prime$ resonance, that in any case modifies just slightly the results.}
\label{tab:highdim}
\end{table}

Our results agree with those of Ref.~\cite{Cirigliano:2002jy,Cirigliano:2003kc,Dominguez:2003dr,Bordes:2005wv} since they also use pinched weights, but we think ours are based on much more solid grounds, due to the completely different approach followed. We see in fact that the DV error was underestimated in Refs.~\cite{Cirigliano:2002jy,Cirigliano:2003kc}, especially in the determination of the higher-dimensional condensates\footnote{This is just an explicit case where we can see that even when the pinched weights generate less DV than the standard weights $s^n$, the observed plateau is in part artificially created and hides the DV. That is why the errors of Ref.~\cite{Cirigliano:2003kc} are underestimated.}.

We also agree with the results of Ref.~\cite{Bijnens:2001ps} based on the use of the second duality point, although that technique has a much larger error. It is remarkable also the agreement with Ref.~\cite{Rojo:2004iq} that is the only one that follows a technique similar to ours, trying to analyze the possible behavior of the spectral function but through a neural-network approach. Their result has a bigger uncertainty, maybe only due to the fact they used the old ALEPH data.

Our analysis indicates that the DV error associated to the use of the first duality point is very large and was grossly underestimated in Ref.~\cite{Zyablyuk:2004iu,Narison:2004vz}, where also higher-dimensional condensates were neglected. In Ref.~\cite{Friot:2004ba,Masjuan:2007ay,Cata:2009fd} the numerical values obtained at
this first duality point are supported through theoretical analyses based on the so-called ``minimal hadronic ansatz'' (a large-$N_C$-inspired 3-pole model) or Pad\`e approximants. Our results show however that the first duality point is very unstable when we change from the WSRs to the $\cO_{6,\ldots,16}$ sum rules, indicating that the systematic error of these approaches is non-negligible. Essentially the same can be said about Refs.~\cite{Barate:1998uf,Davier:1998dz} where the last available point $s_0=m_\tau^2$ was used.
The minimal hadronic ansatz \cite{Friot:2004ba,Cata:2009fd} gives a reasonable approximation to $\mathcal{O}_6$, but its accuracy seems not good enough to reproduce the signs of the higher-order condensates (although an alternating-sign series is indeed predicted).

Summarizing, our results agree within two sigmas with the other estimates of $\cO_6$, but for condensates of higher dimension $\cO_{8,10,12,14,16}$ they agree with Refs.~\cite{Cirigliano:2002jy,Cirigliano:2003kc,Dominguez:2003dr,Bordes:2005wv,Bijnens:2001ps,Rojo:2004iq} but not with Refs.~\cite{Barate:1998uf,Davier:1998dz,Zyablyuk:2004iu,Narison:2004vz,Peris:2000tw,Friot:2004ba}. It is worth noting that in particular our method shows that $\cO_6$ and $\cO_8$ are both negative, whereas it suggests that the sign alternates for higher-dimensional condensates.

\section*{Acknowledgments}
This work has been supported in part by the EU MRTN network FLAVIAnet [Contract No. MRTN-CT-2006-035482], by MICINN, Spain 
[Grants FPA2007-60323 (M.G.-A., A.P), FPA2006-05294 (J.P.) and Consolider-Ingenio 2010 Program CSD2007-00042 --CPAN--], by Generalitat Valenciana [Prometeo/2008/069 (A.P.)] and by Junta de Andaluc\'{\i}a (J.P.) [Grants P07-FQM 03048 and P08-FQM 101]. The work of M.G.-A. is funded through an FPU Grant (MICINN, Spain).

\providecommand{\href}[2]{#2}\begingroup\raggedright\endgroup

\end{document}